\documentclass[12pt]{article}
\usepackage[hypertex]{hyperref}
\usepackage[dvipdfmx]{graphicx} 
\usepackage{graphicx,amsmath,amssymb,amsfonts,cite,bm,cancel,colortbl}

\def\a{\alpha} \def\b{\beta} \def\g{\gamma}  \def\d{\delta} \def\D{\Delta} \def\e{\epsilon} \def\z{\zeta}     \def\l{\lambda} \def\L{\Lambda}  \def\n{\nu}      \def\s{\sigma} \def\S{\Sigma} \def\t{\tau} \def\ph{\phi} \def\Ph{\Phi}     

\def\dg{\dagger} \def\del{\partial} \def\nn{\nonumber}

 \newcommand{\eV}{ {\rm eV} }   \newcommand{\GeV}{ {\rm GeV} } 

\def\ds{\displaystyle}     

\newcommand{\sla}[1]{#1\!\!\!\!/ \,}

\newcommand{\lsp}{ \left ( } \newcommand{\rsp}{ \right ) } \newcommand{\Lg}{\mathcal{L}}  \newcommand{\To}{\Rightarrow}   

\newcommand{\vev}[1]{ \langle {#1} \rangle }

\newcommand{\row}[2]{ \begin{pmatrix}  #1 & #2   \end{pmatrix}  }
\newcommand{\column}[2]{ \begin{pmatrix}  #1 \\ #2 \\  \end{pmatrix} }
\newcommand{\Row}[3]{ \begin{pmatrix} #1 & #2 & #3 \end{pmatrix} }
\newcommand{\Column}[3]{ \begin{pmatrix} #1 \\ #2 \\ #3 \end{pmatrix} }
\newcommand{\diag}[2]{ \begin{pmatrix}  #1 & 0 \\ 0 & #2 \\   \end{pmatrix}  }
\newcommand{\offdiag}[2]{ \begin{pmatrix} 0 & #1 \\ #2 & 0 \\   \end{pmatrix} }
\newcommand{\Diag}[3]{ \begin{pmatrix} #1 & 0 & 0 \\ 0 & #2 & 0 \\ 0 & 0 & #3 \\\end{pmatrix}}

\usepackage{color}

\begin{document}
\begin{flushright}
STUPP-18-235
\end{flushright}

\vskip 1.35cm

\begin{center}
{\large \bf Lopsided texture compatible with thermal leptogenesis \\ in partially composite Pati--Salam unification}

\vskip 1.2cm

Masaki J. S. Yang

\vskip 0.4cm

{\it Department of Physics, Saitama University, \\
Shimo-okubo, Sakura-ku, Saitama, 338-8570, Japan\\
}

\begin{abstract} 

In this paper, we consider a lopsided flavor texture compatible with thermal leptogenesis 
in partially composite Pati--Salam unification. 
The Davidson--Ibarra bound $M_{\nu R1} \gtrsim 10^9 \GeV$ for the successful thermal leptogenesis can be recast to the Froggatt--Nielsen (FN) charge of the lopsided texture. 
We found the FN charge $n_{\nu1}$ of the lightest right-handed neutrino $\nu_{R1}$ can not be larger than a upper bound, $n_{\nu1} \lesssim 4.5$.

From the viewpoint of unification, the FN charges of the neutrinos $n_{\nu i}$ 
should be the same to that of other SM fermions. Then, two cases $n_{\nu i} = n_{qi} = (3,2,0)$ and $ n_{\nu i} = n_{l i} = (n+1,n,n)$ are considered. 
Observations of PS model shows that the case of $n=0$, 
$n_{li} = n_{di} = (1,0,0)$ will be the simplest realization. 

To decrease the FN charges of these fermions from the GUT invariant FN charges
$n_{qi} = (3,2,0)$, we utilize the partial compositeness. 
In this picture, the hierarchies of Yukawa matrices are a consequence of 
mixings between massless chiral fermions $f_{L}, f'_{R}$ and massive vector fermions $F_{L,R}, F'_{L,R}$. 
This is induced by the linear mixing terms $\lambda^{f} \bar f_{L} F_{R}$ and $\lambda^{f'} \bar F'_{L} f'_{R}$. 

As a result of the partial compositeness, 
the decreases of FN charges require
fine-tunings between mass and Yukawa matrices 
either for the increases of $\lambda^{f, f'}$ or for the decreases of $M_{F,F'}$. 
Therefore, the case for $n=2$ and $n_{di} = n_{li} = (3,2,2)$, 
which requires only increases of FN charges 
will be appropriate to build a natural model. 

Moreover, it is found that 
composite neutrino sector should have (almost) the same flavor structure
 to reproduce the large mixing of neutrinos by the type-I seesaw mechanism.
If the vev of GUT breaking Higgs mediates flavor structure, 
they contribute to some mass term.  
Then, this statement  can be hold for even in other Pati--Salam model, 
that does not assume the partial compositeness.  

\clearpage

\end{abstract} 

\end{center}



\section{Introduction}

The peculiar flavor structure of the Standard Model (SM) 
can be some hints of the theoretical origin of the Higgs boson and the flavor puzzle. 
On the other hand, analysis of flavor structures strongly depend on how we consider the origin of the Higgs boson. 
Then, study of model independent flavor textures is one of the dominant approach.  
For example, the Fritzsch texture \cite{Fritzsch:1977vd, Fritzsch:1995nx, Fritzsch:1999ee, Nishiura:1999yt,Matsuda:1999yx, Xing:2003zd, Bando:2003wb,Bando:2004hi}, 
the democratic texture \cite{Harari:1978yi, Koide:1983qe, Koide:1989zt, Tanimoto:1989qh, Fritzsch:1989qm, Babu:1990fr, Fritzsch:1995dj, Fukugita:1998vn, Fukugita:1998kt, Fujii:2002jw, Fritzsch:2004xc, Xing:2010iu, Zhou:2011nu, Yang:2016esx}, 
and the lopsided texture \cite{Sato:1997hv, Nomura:1998gm, Bando:2000gs, King:2000ge, Maekawa:2001uk, Asaka:2003fp, Kitano:2003cn}. 

Among them, 
the lopsided texture appears to be more natural by two reasons:
\begin{itemize}
\item If we assume the type-I seesaw mechanism \cite{seesaw}, 
 majorana mass matrix of right-handed neutrinos would be waterfall texture whether Yukawa matrix is cascade or waterfall \cite{Yang:2016crz} in Table 1.  
Then, waterfall is more desirable for the unified description of flavor.

\item If a waterfall texture is symmetric matrix, 
quark Yukawa matrices should have approximate zero texture \cite{Hall:1993ni} in order to realize the CKM matrix \cite{Cabibbo:1963yz, Kobayashi:1973fv}. 
In some sense, zero textures in low energy is unnatural without a complicated symmetry. 
Then, the asymmetric waterfall texture appears to be more natural. 
\end{itemize}
\begin{table} [ht]
\centering
\begin{math}
\begin{array}{cc}
\begin{array}{|c|c|}
\hline
\begin{pmatrix}
\e & \e & \e \\
\e & \d & \d \\
\e & \d & 1 \\
\end{pmatrix}
&
\begin{pmatrix}
\e^{2} & \e \d & \e  \\
\e \d  & \d^{2} & \d \\
\e & \d & 1 \\
\end{pmatrix}
\\[7pt]
\hline
\text{Cascade} & \text{Waterfall} \\ \hline
\end{array}
& ~~~
Y_{u} = 
\begin{pmatrix}
\ll m_{u} & \sqrt{m_{u} m_{c}} & \ll \sqrt{m_{u} m_{t}} \\
\sqrt{m_{u} m_{c} } & m_{c} & \ll \sqrt{m_{c} m_{t}} \\
\ll \sqrt{m_{u} m_{t}} &  \ll \sqrt{m_{c} m_{t}} & m_{t}
\end{pmatrix}
\\ \\
(a) & (b)
\end{array}
\end{math}
\caption{(a) The cascade and waterfall texture for $1 \gg \d \gg \e$ \cite{Haba:2008dp}. (b) An example of symmetric Yukawa texture compatible with CKM matrix \cite{Hall:1993ni}.}
\label{obs}
\end{table}

In this paper, we consider a lopsided flavor texture compatible with thermal leptogenesis 
in partially composite Pati--Salam unification. 
The Davidson--Ibarra (DI) bound \cite{Davidson:2002qv, Hamaguchi:2001gw} for the successful thermal leptogenesis can be recast to the Froggatt--Nielsen (FN) charge \cite{Froggatt:1978nt} of the lopsided texture. 
We found the FN charge $n_{\n1}$ of the lightest right-handed neutrino $\n_{R1}$ can not be larger than a upper bound, $n_{\n1} \lesssim 4.5$.

The Grand Unified Theory (GUT) \cite{Georgi:1974sy} is suitable 
to explore the unified origin of flavor structures.
Since the proton decay have not been observed for a long time \cite{Miura:2016krn},  
it is somewhat reasonable to consider the Pati--Salam unification \cite{Pati:1974yy},
 a GUT model with no proton decay. 
From the viewpoint of unification, the FN charges of the neutrinos $n_{\n i}$ 
should be the same to that of other SM fermions. 
Then, two cases $n_{\n i} = n_{qi} = (3,2,0)$ and $ n_{\n i} = n_{l i} = (n+1,n,n)$ are considered. 
Observations of PS model shows that the case of $n=0$, 
$n_{li} = n_{di} = (1,0,0)$ will be the simplest realization. 

To decrease the FN charges of these fermions from the GUT invariant FN charges
$n_{qi} = (3,2,0)$, we utilize the partial compositeness \cite{Kaplan:1991dc, Contino:2006nn}. 
In this picture, the hierarchies of Yukawa matrices are a consequence of 
mixings between massless chiral fermions $f_{L}, f'_{R}$ and massive vector fermions $F_{L,R}, F'_{L,R}$. 
This is induced by the linear mixing terms $\l^{f} \bar f_{L} F_{R}$ and $\l^{f'} \bar F'_{L} f'_{R}$. 

If the GUT breaking Higgs contributes these linear mixing terms, 
the resulting Yukawa interactions can be different between quarks and leptons.
For this purpose, we employ the bi-fundamental Higgs $H_{R} $({\bf 4,1,2}) under the PS group $G_{PS} = SU(4)_{c} \times SU(2)_{L} \times SU(2)_{R}$ \cite{Antoniadis:1988cm}. 
In order to mediate the GUT breaking effect to the quarks and leptons, exotic fermions $\psi_{L}^{\a\b} ({\bf 6,1,1}) \sim (\tilde D^{c}_{R}, \tilde D_{L})$ and $\chi_{R}^{ab} ({\bf 1,2,2}) \sim (\tilde L^{c}_{L}, \tilde L_{R})$ are introduced. They can form a $\bf 10$ representation in $SO(10)$. 
Yukawa interactions between exotics and composites generate GUT breaking linear mixings. 

The lopsided texture in partially composite models 
is generated from the particular set of mass terms $\l^{f_{L}} M_{F}^{-1}$ and $M_{F'}^{-1} \l^{f_{R}}$. Then, decreases of FN charges are realized by 
increases of $\l^{f_{L}, f_{R}}$ or decreases of $M_{F, F'}$. 
Since the couplings between exotics and elementals are forbidden, 
the increases of $\l^{f_{L}, f_{R}}$ from mixings are difficult without fine-tunings for $M_{F, F'}, Y^{6} V \gg \l^{f_{L}, f_{R}}$. 
On the other hand, the decreases of $M_{F,F'}$ require an order of $\l^{2} \simeq 5 \%$ fine-tunings between matrices $Y^{6} V, \tilde Y^{6} V$ and $M_{F, F'}$. 

As a result of the partial compositeness, 
the decreases of FN charges require
fine-tunings between mass and Yukawa matrices 
either for the increases of $\l^{f, f'}$ or for the decreases of $M_{F,F'}$. 
Therefore, the case for $n=2$ and $n_{di} = n_{li} = (3,2,2)$, 
which requires only increases of FN charges 
will be appropriate to build a natural model. 

Moreover, it is found that 
composite neutrino sector should have (almost) the same flavor structure
 to reproduce the large mixing of neutrinos by the type-I seesaw mechanism.
If the vev of GUT breaking Higgs mediates flavor structure, 
they contribute to some mass term.  
Then, this statement  can be hold for even in other Pati--Salam model, 
that does not assume the partial compositeness.  

This paper is organized as follows. 
In the next section, we review the lopsided texture and the DI bound. 
In Sec.~3, the lopsided texture in PS GUT is overviewed. 
In Sec.~4, the partial compositeness is reviewed. 
In Sec.~5, we consider a partially composite Pati--Salam Unification. 
The final section is devoted to conclusions and discussion.

\section{Thermal Leptogenesis with Lopsided Texture}

In this section, we discuss how the Davidson--Ibarra bound of the thermal leptogenesis \cite{Davidson:2002qv, Hamaguchi:2001gw}
restricts the Froggatt--Nielsen (FN) charge \cite{Froggatt:1978nt} of the lopsided texture. 
First of all, the Yukawa interactions of the SM is defined as
\begin{align}
\Lg \ni  \sum_{f} - y_{f ij} \bar f_{Li } f'_{Rj} H + {\rm h.c.} \, ,
\end{align}
for the SM fermions $f=q,l, f' = u,d,\n,e$ and the Higgs boson $H$.  The lopsided texture (at  the GUT scale $\L_{\rm GUT} \simeq 2 \times 10^{16} \, \GeV$) 
is represented as 
\begin{align}
y_{u} \propto
\begin{pmatrix}
\l^{6} & \l^{5} & \l^{3} \\
\l^{5} & \l^{4} & \l^{2} \\
\l^{3} & \l^{2} & \l^{0}
\end{pmatrix}
, ~~~
y_{d} \propto y_{e}^{T} \propto  
\begin{pmatrix}
\l^{4} & \l^{3} & \l^{3} \\
\l^{3} & \l^{2} & \l^{2} \\
\l^{1} & \l^{0} & \l^{0}
\end{pmatrix}
, ~~~ 
m_{\n} \propto 
\begin{pmatrix}
\l^{2} & \l^{1} & \l^{1} \\
\l^{1} & \l^{0} & \l^{0} \\
\l^{1} & \l^{0} & \l^{0}
\end{pmatrix} .
\end{align}
Here, $\l$ is the Cabibbo angle $\l \simeq 0.22$.
If the light neutrino mass is induced by the type-I seesaw mechanism \cite{seesaw}, 
the neutrino Yukawa and  heavy majorana mass matrices should have the following form
\begin{align}
y_{\n} \propto 
\begin{pmatrix}
\l^{n_{\n1} + 1} & \l^{n_{\n2} + 1} & \l^{n_{\n3} + 1} \\
\l^{n_{\n1}} & \l^{n_{\n2}} & \l^{n_{\n3}} \\
\l^{n_{\n1}} & \l^{n_{\n2}} & \l^{n_{\n3}}
\end{pmatrix}
, ~~~ 
M_{\n R} \propto 
\begin{pmatrix}
\l^{2 n_{\n1}} & \l^{n_{\n1} + n_{\n2}} & \l^{n_{\n1} + n_{\n3}} \\
\l^{n_{\n1} + n_{\n2}} & \l^{2 n_{\n2}} & \l^{n_{\n2} + n_{\n3}} \\
\l^{n_{\n1} + n_{\n3}} & \l^{n_{\n2} + n_{\n3}} & \l^{2 n_{\n3}} \\
\end{pmatrix} ,
\end{align}
in order to realize the large mixing of MNS matrix \cite{Maki:1962mu}. 
These textures, realized by the $U(1)$ Froggatt--Nielsen (FN) charges in Table 2,
appears to be more natural by two reasons, as we mentioned at the introduction:
\begin{itemize}
\item If we assume the type-I seesaw mechanism, 
 majorana mass matrix would be waterfall texture whether Yukawa matrix is cascade or waterfall \cite{Yang:2016crz} in Table 1.  
Then, waterfall is more desirable for the unified description of flavor.

\item If a waterfall texture is symmetric matrix, 
quark Yukawa matrices should have approximate zero texture \cite{Hall:1993ni} in order to realize 
CKM matrix \cite{Cabibbo:1963yz, Kobayashi:1973fv}. In some sense, zero texture in low energy is unnatural without a complicated symmetry. Then, the asymmetric waterfall texture appears to be more natural. 
\end{itemize}
\begin{table}[htb]
  \begin{center}
    \begin{tabular}{|c|ccc|ccc|ccc|} \hline
        Field   & ${\bf 10}_{1}$ & ${\bf 10}_{2}$ & ${\bf 10}_{3}$ & $\bar{\bf 5}_{1}$ & $\bar{\bf 5}_{2}$ & $\bar{\bf 5}_{3}$  & ${\bf 1}_{1}$ & ${\bf 1}_{2}$ & ${\bf 1}_{3}$ \\ \hline 
      $U(1)$ &  3 & 2 & 0 & $n+1$ & $n$ & $n$ & $n_{\n 1}$ & $n_{\n 2}$ & $n_{\n 3}$ \\ \hline
    \end{tabular}
    \caption{The FN charge assignments of the SM fermions grouped into the representations of $SU(5)$, 
    ${\bf 10}_{i} = (q_{L}, u_{R}^{c}, e_{R}^{c})_{i} , \bar{\bf 5}_{i} = (d_{R}^{c}, l_{L})_{i} , {\bf 1}_{i} = \n_{R i}^{c}.$}
  \end{center}
\end{table}

The thermal leptogenesis with the lopsided texture have been discussed 
in several papers \cite{Asaka:2003fp, Cannoni:2013gq, Ishihara:2015uua}.
They agreed that larger $n_{\n i}$ are incompatible with the DI bound.
Here let us confirm this fact systematically. 

\vspace{12pt}

In order to retain the room for adjustment of FN charges, the 
two Higgs doublet model (2HDM) is assumed.
The fermion mass matrices are given by
\begin{eqnarray}
m_{u ij} = {v\over\sqrt{2}} y_{u ij}  s_{\b} , & 
m_{d ij} = \ds {v\over\sqrt{2}} y_{d ij}  c_{\b} , \\
m_{\nu ij}^{\rm Dirac} = {v\over\sqrt{2}} y_{\n ij}  s_{\b} , &
 m_{e ij} = \ds {v\over\sqrt{2}}  y_{e ij}  c_{\b} , 
\end{eqnarray}
where $\tan \b \equiv {v_{u} / v_{d}}, \, v_{u} = v \sin \b \equiv v s_{\b}, v_{d} = v \cos \b \equiv v c_{\b} $. 

In this case FN charges of leptons $n_{l i}, n_{e i}, n_{\n i}$ have dependence of $\tan \b$ through the mass relations, 
\begin{align}
m_{\n i}^{\rm Dirac} \simeq {1\over\sqrt{2}} \l^{n_{li} + n_{\n i}} v c_{\b} , ~~~~
 m_{e i} \simeq {1\over\sqrt{2}} \l^{n_{li} + n_{ei}} v c_{\b} .
\end{align}
Tentatively  $n_{l i}, n_{e i}, n_{\n i}$ are treated 
as free parameters, 
without fixing them like in Table 2.
The light neutrino mass is given by
\begin{align}
m_{\n} \equiv m 
\begin{pmatrix}
\l^{2 n_{l1}} & \l^{n_{l1} + n_{l 2}} & \l^{n_{l1} + n_{l 3}}  \\
 \l^{n_{l1} + n_{l 2}} & \l^{2 n_{l2}} &  \l^{n_{l2} + n_{l 3}} \\
 \l^{n_{l1} + n_{l 3}} & \l^{n_{l2} + n_{l3}} &  \l^{2 n_{l 3}} \\
\end{pmatrix} .
\end{align}
In the many model with lopsided textures, 
the mass eigenvalues of the lighter neutrinos $m_{\n i}$ 
are roughly fixed as
\begin{align}
m_{\n}^{\rm diag} 
\sim m \Diag{\l^{2 n_{l1}}}{\l^{2 n_{l2}}}{\l^{2 n_{l3}}}
\sim \Diag{0.002}{0.01}{0.05} [\eV] .
\end{align}
Then, the overall factor $m \sim \l^{- 2 n_{l1}} 0.002 \, [\eV] \sim \l^{-2 n_{l3}} 0.05 \, [\eV]$ also depends on $\tan \b$ through the FN charge of the left-handed leptons $n_{li}$.

By the seesaw mechanism, the heavy majorana mass matrix can be reconstructed as follows
\begin{align} 
&M_{\n R} = {v^{2} s_{\b}^{2} \over 2} y_{\n}^{T} m_{\n}^{-1} y_{\n}, \\
& = {v^{2} s_{\b}^{2} \over 2 m} 
\begin{pmatrix}
\l^{2 n_{\n1}} & \l^{n_{\n1} + n_{\n2}} & \l^{n_{\n1} + n_{\n3}} \\
\l^{n_{\n1} + n_{\n2}} & \l^{2 n_{\n2}} & \l^{n_{\n2} + n_{\n3}} \\
\l^{n_{\n1} + n_{\n3}} & \l^{n_{\n2} + n_{\n3}} & \l^{2 n_{\n3}} \\
\end{pmatrix}
\sim 
{246 ^{2} \, [\GeV^{2}] s_{\b}^{2} \l^{2n_{\n1}} \over 2m} 
\Diag{1}{\l^{2 (n_{\n2} - n_{\n1})}}{\l^{2(n_{\n3} - n_{\n1})}} .
\end{align}
If we assume the normal hierarchy $n_{\n 1} > n_{\n 2} > n_{\n 3}$ for $\n_{Ri}$, 
the FN charge of the lightest right-handed neutrino $n_{\n 1}$ is bounded by the DI bound:
\begin{align}
M_{\n R 1} \sim {6 \times 10^{4} \, [\GeV^{2}] s_{\b}^{2} \l^{2n_{\n 1}}  \over 2 \l^{- 2 n_{l1}} 0.002 \, [\eV] }
& \gtrsim 10^{9} \, [\GeV] , \\
 6 \times 10^{14} \, [\GeV] s_{\b}^{2} \l^{2 (n_{\n1} + n_{l1} - 1)} & \gtrsim 10^{9} \, [\GeV] , \\
1.5 \times 10^{7}  \l^{2 (n_{\n1} + n_{l1})} & \gtrsim  1 ,  ~~ \To ~~
5.5 \gtrsim (n_{\n1} + n_{l1} ) .
\label{FNbound}
\end{align}
The factor  $6 \times 10^{14}$ is well-known result of the seesaw scale \cite{Babu:1998wi}. 
In the third line, we set $s_{\b} \simeq 1$. 
The result shows that successful thermal leptogenesis requires larger $\tan \b$ and 
smaller FN charges, in the model with lopsided textures. 
For example, $n_{\n1} \lesssim 4.5$ for $n_{l1} = 1$ (large $\tan \b$),  and $n_{\n1} \lesssim 2.5$ for $n_{l1} = 3$ ($\tan \b \sim 1$). 
This result is consistent with the previous papers \cite{Cannoni:2013gq, Ishihara:2015uua}.

However, this bound can be applied only for the 
strongly hierarchical right-handed neutrinos $M_{2,3} > 100 M_{1}$ \cite{Hambye:2003rt, Davidson:2003yk}.
It corresponds to the case $n_{\n 1} \gtrsim n_{\n 2} + 1.5$, where 
the hierarchy of $M_{\n R}$ is about as same as that of up-type quarks. 

\section{Lopsided texture in Pati--Salam GUT: overview}

The Grand Unified Theory (GUT) \cite{Georgi:1974sy} is suitable 
to explore the unified origin of flavor structures. 
The lopsided texture have been embedded to SO(10) \cite{Nomura:1998gm, Asaka:2003fp, Kitano:2003cn}, $E_{6}$ \cite{Bando:2000gs, Maekawa:2001uk}, and originally $E_{7}$ \cite{Sato:1997hv}. 
These GUTs were well researched because they predict proton decay. 
One of the latest bound of the proton decay is $\t / B(p \to e^{+} \pi^{0}) > 1.6 \times 10^{34}$ years at 90\% confidence level \cite{Miura:2016krn}. 
In contrast, 
the Pati--Salam (PS) unified model \cite{Pati:1974yy} has no proton decay, unless the model has $f_{i}^{c} [y \phi_{ij} + \tilde y \e^{ijkl} \ph_{kl}] f_{j}$ type coupling with a 6 representation scalar $\ph_{ij}$ under $SU(4)_{c}$ \cite{Li:1982hu}. 
Since a proton decay have not been observed for a long time,  
it is somewhat reasonable to consider a GUT model without proton decay. 
The lopsided texture in Pati--Salam unification 
is also realized by FN mechanism \cite{King:2000ge}.
\vspace{12pt}

First of all, let us discuss on the relation between FN charges and the DI bound 
in the PS model. 
From the viewpoint of unification, the FN charges of the neutrinos $n_{\n i}$ 
should be the same to that of other SM fermions. 
Then, two cases $n_{\n i} = n_{qi} = (3,2,0)$ (here we name this ``quark type'') and $ n_{\n i} = n_{l i} = (n+1,n,n)$ (``lepton type'') are considered. 
Smaller (larger) $n$ charge corresponds to large (small) $\tan \b$. 
Here we will constrain the FN charges by several physical suggestions,  
presented as Table 3. 

{\bf $\bullet$ Representation of GUT Higgs field:}
In order to generate different flavor structures in the PS model, 
GUT breaking Higgs should mediate flavor dependence in some way.
Usually the symmetry breaking of the PS model is achieved by the following two Higgs fields
$\S ({\bf 15,1,1})$, $\D_{R} ({\bf 10,1,3})$ under the group $G_{PS} \equiv SU(4)_{c} \times SU(2)_{L} \times SU(2)_{R}$:
\begin{align}
\vev{\S} = V 
\begin{pmatrix}
 1 & 0 & 0 & 0 \\ 0 & 1 & 0 & 0 \\ 0 & 0 & 1 &0 \\ 0 & 0 &0 & -3 \\
\end{pmatrix} \otimes \diag{1}{1},  ~~~ 
\vev{\D_{R}} = V' 
\begin{pmatrix}
 0 & 0 & 0 & 0 \\ 0 & 0 & 0 & 0 \\ 0 & 0 & 0 &0 \\ 0 & 0 &0 & 1 \\
\end{pmatrix}
\otimes \offdiag{0}{1} .
\end{align}
However, it is difficult to induce different flavor structures between quarks and leptons, 
unless an exponential couplings such as $e^{- y \S / \L}$ are assumed. 

Alternatively, we employ a bi-fundamental representation
\begin{align}
H_{R} \, ({\bf 4,1,2}) = (u_{RH}, d_{RH}, \n_{RH}, e_{RH}), 
\end{align}
under $G_{PS}$ \cite{Antoniadis:1988cm, King:2000ge}. 
This Higgs corresponds  $\bf 16$ representation of SO(10) and the truly minimal Higgs \cite{Brahmachari:2003wv} 
in the left-right symmetric model. 
$H_{R}$ breaks the group $G_{PS}$ to SM by 
obtaining a vacuum expectation value (vev) in the  ``right-handed neutrino'' direction:
\begin{align}
\vev{H_{R}} = \vev{\n_{RH}}  
\sim 10^{16} \GeV ,
\label{higgsvev}
\end{align}
and then the breaking scale is determined uniquely.
Therefore, GUT invariant FN charges are determined as $n_{f i} = (3,2,0)$, 
because the vev of $H_{R}$ is difficult to couple quarks. 

{\bf $\bullet$ Davidson--Ibarra bound:} Since the FN charges of the lightest right-handed neutrino have the upper bound, 
thermal leptogenesis fails for the quark type with small  $\tan \b$.
There is two ways to avoid this DI bound. 
The one way is to satisfy the bound $n_{\n1} \lesssim 4.5$. 
The other is to weaken the hierarchy of the mass of right-handed neutrinos, 
because  the DI bound does not holds 
for mild-hierarchical $\n_{R i}$, in such a case of $M_{\n R} \propto m_{\n}$ \cite{Hambye:2003rt, Davidson:2003yk}.

{\bf $\bullet$ Magnitude correlation between $M_{\rm GUT}$ 
and $M_{F}$:} 
Here, $M_{F}$ is the scale where the flavor structure is produced. 
If $M_{F} \gg M_{\rm GUT}$ holds, 
GUT breaking flavor effects stay within perturbative range. 
Then, this case can induce only increases of FN charges, such as $(3,2,0) \to (3,2,2)$. 
On the other hand, decreases of the FN charges $(3,2,0) \to (1,0,0)$
imply that the GUT scale is larger than the scale $M_{F}$. 
Here, the case of $n=1$ (or $(3,2,0) \to (2,1,1)$) is not considered, because it requires 
both increases and decreases of FN charges.
These discussion is summarized in Table 3. 

The quark type appears to be the simplest realization, 
because the latter case requires additional change of the flavor structure of neutrinos, 
$n_{\n i} = (3,2,0) \to (n+1,n,n)$. 
We  will discuss the latter possibility in the next paper, 
and focus on the former one. 
In this case, the FN charges of leptons (and down-type quarks) should decrease $n_{li} = (3,2,0) \to (1,0,0)$ by the vev of GUT Higgs Eq.~(\ref{higgsvev}). 
In the next section, we will discuss the partial compositeness, 
because this theory can decrease the FN charges. 

\begin{table}
\begin{center}
\begin{math}
\begin{array}{|c|c|c|}
\hline
 & \text{quark type} & \text{lepton type} \\ \hline
 & \cellcolor[gray]{0.99} n_{\n i} = (3,2,0) & \cellcolor[gray]{0.9} n_{\n i} = (1, 0,0) \\ 
n = 0  & \cellcolor[gray]{0.99} n_{l i} = (1,0,0) & \cellcolor[gray]{0.9} n_{l i} = (1,0,0)  \\ 
\tan \b \simeq 40 & \cellcolor[gray]{0.99} \text{it will be the} & \cellcolor[gray]{0.9} \text{it requires another } \\ 
  M_{\rm GUT} \gtrsim M_{F}  & \cellcolor[gray]{0.99}  \text{simplest realization} & \cellcolor[gray]{0.9} \text{ change of $n_{\n i}$ } \\ \hline
 & \cellcolor[gray]{0.66} n_{\n i} = (3,2,0) & \cellcolor[gray]{0.9} n_{\n i} = (3, 2, 2) \\ 
n = 2 & \cellcolor[gray]{0.66} n_{l i} = (3,2,2) & \cellcolor[gray]{0.9} n_{l i} = (3,2,2)   \\ 
 \tan \b \simeq 1  & \cellcolor[gray]{0.66} \text{ incompatible with } & \cellcolor[gray]{0.9} \text{ it requires another } \\ 
  M_{\rm GUT} \ll M_{F} & \cellcolor[gray]{0.66} \text{ thermal leptogenesis} & \cellcolor[gray]{0.9} \text{ change of $n_{\n i }$} \\ \hline
\end{array}
\end{math}
\caption{Relation between the FN charges, leptogenesis, the flavor scale $M_{F}$ and the GUT scale $M_{\rm GUT}$.}
\end{center}
\end{table}
%

\section{Partial Compositeness in Composite Higgs Model}

Construction of the lopsided texture is basically classified into two ways: FN mechanism, and mixing between SM and heavy fermions\footnote{Composite models called ``dual FN mechanism'' are also considered \cite{Kaplan:1997tu, Haba:1998wf}.}. 
The typical example of the latter case is $E_{6}$ twist mechanism \cite{Bando:2000gs}, universal seesaw \cite{Berezhiani:1983hm,Berezhiani:1985in}, partial compositeness \cite{Kaplan:1991dc, Contino:2006nn}, and so on.
Among them, a paper with the $E_{6}$ twist naively have failed the thermal leptogenesis \cite{Ishihara:2015uua}, 
because the model should have large FN charge $n_{\n i} = (3,2,0)$ and small $\tan \b$.  
Here, we consider the partial compositeness for realization of the lopsided texture. 

The basic idea of the partial compositeness is that the SM fields at low energy are the mixed states between elemental (massless) fields and composite (massive) fields, like $\rho - \g$ mixing. 
Flavor structures are induced from mixings between massive and massless fermions with the same quantum numbers. 
In this section, we will shortly review the composite Higgs model with partial compositeness.

\subsection{Partial compositeness}

The original context basically assumes lower composite scale around TeV 
and some large global symmetry which contains the SM gauge group. 
The minimal model has the strong sector with a global symmetry $SO(5) \times U(1)_{X}$ which is broken down to $SO(4) \times U(1)_{X}$ at the scale $f$ \cite{Agashe:2004rs, Contino:2006qr}.
In this paper, we do not assume the compositeness and such a symmetry. 

The composite Higgs model (under the breaking scale $f$) can be described by a simplified two-site description \cite{Contino:2006nn}, where the composite sector is replaced by the first resonances which mix with the SM fields. 
The Lagrangian is divided to three parts: 
\begin{equation}
\Lg = \Lg_{\rm composite} + \Lg_{\rm elementary} + \Lg_{\rm mixing} .
\end{equation}
The linear mixing terms $\Lg_{\rm mixing}$ represent mass terms between massive  and massless fields. 
Due to this mixing, massless eigenstates which are identified with the SM fields are superposition of elementary and composite states. 
The Lagrangian for these fermions are written as
\begin{align}
\Lg_{\rm composite}& = \bar{Q}_i (i \sla{D} -M_{Q i}) Q_i + \bar{U}_i (i \sla{ D} -M_{U i}) U_i + \bar{D}_i (i \sla{D}-M_{D i}) D_{i} \nn \\
& + \bar{L}_i (i \sla{D} - M_{L i}) L_i + \bar{N}_i (i \sla{ D} - M_{N i}) N_i + \bar{E}_i (i \sla{ D} - M_{E i}) E_{i} \label{Lcomp0} \\
& +Y^U_{ij} \bar{Q}_{Li} \widetilde{H} U_{Rj}+ Y^D_{ij} \bar{Q}_{Li} H D_{Rj} 
+ Y^{N}_{ij} \bar L_{Li} \widetilde{H} N_{Rj} + Y^{E}_{ij} \bar L_{Li} H E_{Rj} \nn \\
&+\widetilde{Y}^U_{ij} \bar{U}_{L i} H Q_{R j}+ \widetilde{Y}^D_{ij} \bar{D}_{L i} \widetilde H Q_{R j} + \widetilde{Y}^N_{ij} \bar{N}_{L i} H L_{R j}+ \widetilde{Y}^E_{ij} \bar{E}_{L i} \widetilde H L_{R j}+ {\rm h.c.}
\, , \label{Lcomp} \\
\Lg_{\rm elementary} & =  i \bar q _{L i} \sla D q_{L i} + i \bar u_{Ri} \sla D u_{Ri} + i \bar d_{Ri} \sla D d_{Ri} + i \bar l_{Li} \sla D l_{Li} + i \bar \n_{Ri} \sla D \n_{Ri} + i \bar e_{Ri} \sla D e_{Ri} \, ,  \\
\Lg_{\rm mixing} &= \l^{q}_{ij} \bar{q}_{Li} Q_{Rj} + \l^{u}_{ij} \bar{U}_{Li} u_{Rj} + \l^{d}_{ij} \bar{D}_{Li} d_{Rj} 
+ \l^{l}_{ij} \bar{l}_{Li} L_{Rj} + \l^{\n}_{ij} \bar{N}_{Li} \n_{Rj} + \l^{e}_{ij} \bar{E}_{Li} e_{Rj} 
+{\rm h.c.} \, .
\label{Lmix}
\end{align}
Here, the small-letter fields $q_{L}, u_{R}, d_{R}, l_{L}, \n_{R}, e_{R}$ are the elemental fermions and the capital-letter fields $Q, U, D, L, N, E$ are vector-like composite fields  with the same gauge charges of corresponding elemental fields. $\tilde H \equiv i \s^{2} H^{*}$ is the conjugate field of the Higgs doublet $H$.

Redefining the physical states, one can obtain the SM Yukawa interactions. 
For the doublet quarks, the mass matrix is rewritten as follows; 
\begin{equation}
\row{\bar q_{Li}}{\bar Q_{Li}}  \column{\lambda^{q}_{ij}}{M_{Qi} \delta_{ij}} Q_{Rj} + {\rm h.c.} \,.
\end{equation}
We can diagonalize this $3 \times 6$ component matrix perturbatively 
for $M_{Qi} \gg \lambda_{ij}^{q}$.
At the leading order, the mass eigenstates of the doublets are found to be
\begin{equation}
\column{q_{Li}^{\rm phys}}{Q_{Li}^{\rm phys}} = 
\begin{pmatrix}
1 - {1\over2} \Delta \Delta^{\dagger} & - \Delta \\
 \Delta^{\dagger} & 1 - {1\over2} \Delta^{\dagger} \Delta \\
\end{pmatrix}_{ij}
\column{q_{Lj}}{Q_{Lj}}
\simeq 
\column{q_{L} - \lambda^{q} M_{Q}^{-1} Q_{L} }{M_{Q}^{-1} \lambda^{q \dagger} q_{L} + Q_{L}}_{i}
+ \mathcal{O} \lsp {\lambda^{q \, 2} \over M_{Q}^{2}} \rsp , 
\label{masseigen}
\end{equation}
where $\Delta_{ij} \equiv \lambda^{q}_{ij} M_{Q j}^{-1}$.
Substituting Eq.~(\ref{masseigen}) and similar equations for other fermions in Eq.~(\ref{Lcomp}), the SM Yukawa interactions are represented by seesaw-like formulae\footnote{We can obtain the same results from integrating out the massive  (composite) fields by solving the equations  
$\partial \Lg / \partial Q = \partial \Lg / \partial U = \cdots =  0$ for the whole Lagrangian. }
\begin{align}
y_{u} = \l^{q} M_{Q}^{-1} Y^{U} M_{U}^{-1} \l^{u},  ~~~ 
y_{\n} = \l^{l} M_{L}^{-1} Y^{N} M_{N}^{-1} \l^{\n},  \\
y_{d} = \l^{q} M_{Q}^{-1} Y^{D} M_{D}^{-1} \l^{d},  ~~~
y_{e} = \l^{l} M_{L}^{-1} Y^{E} M_{E}^{-1} \l^{e} .  
\label{SMYukawas}
\end{align}
In this picture, hierarchies of the SM Yukawa interactions are generated by hierarchies of $\lambda^{f}, M_{F},$ and $Y^{F}$. The terms with $\tilde Y$ do not contribute SM Yukawa matrices in the first order of $\l^{f} / M_{F}$. 
However, the leading approximation can not be valid for the third generation, and we should keep in mind it is just a ``thumb counting''.

\section{Partially Composite Pati--Salam Unification}

Here, we will consider a lopsided flavor texture compatible with thermal leptogenesis in partially composite Pati--Salam unification. 
GUTs with the composite Higgs have been considered in some literatures \cite{Frigerio:2011zg, Nevzorov:2015sha}. PS GUT model were discussed in \cite{Barbieri:2017tuq,Blanke:2018sro}. 
The key observation is that if the vev~(\ref{higgsvev}) induces linear mixing mass terms $\l_{f}$, the difference of quarks and leptons can be generated. 

The field content of the model is shown in Table 4. 
In partially composite models, Higgs boson is a Nambu--Goldstone boson (NGB) of 
some global symmetry and it can only couple with massive composites.  
Here, such a global symmetry is not imposed, and 
we assume that Higgs fields $\Ph, H_{R}$ only couple with massive fermions $F, F'$ and exotics $\psi_{L}, \chi_{R}, \s_{L}$. 
\begin{table}[htb]
  \begin{center}
    \begin{tabular}{|c|ccc|} \hline
           & $SU(4)_{c}$ & $SU(2)_{L}$ & $SU(2)_{R}$ \\ \hline \hline
      $f_{Li}$ & \bf 4 & \bf 2 & \bf 1 \\
      $f_{Ri}$ & \bf 4 & \bf 1 & \bf 2 \\ \hline
      $F_{(L,R)i}$ & \bf 4 & \bf 2 & \bf 1 \\
      $F'_{(L,R)i}$ & \bf 4 & \bf 1 & \bf 2 \\ \hline
      $\psi_{L}$ & \bf 6 & \bf 1 & \bf 1 \\
      $\chi_{R}$ & \bf 1 & \bf 2 & \bf 2 \\ 
      $\s_{L}$  & \bf 1 & \bf 1 & \bf 1 \\ \hline   
      $\Phi$ & \bf 1 & \bf 2 & \bf 2 \\ 
      $H_R$ & \bf 4 & \bf 1 & \bf 2 \\  \hline
    \end{tabular}
    \caption{The charge assignments of the fermions and Higgs fields under the gauge symmetries. The index $i = 1-3$ represents flavor of the fermions.}
  \end{center}
\end{table}

The relevant part of flavor structures in partially composite PS GUT is given by
\begin{align}
\Lg_{\rm composite}& = \bar{F}_{i} (i \sla{D}-M_{F_{i}}) F_{i} + \bar F'_{i} (i \sla{D}-M_{F'_{i}}) F'_{i} +Y^F_{ij} \bar{F}_{Li} {\Ph} F'_{Rj} + {\rm h.c.} \, , \\
%
\Lg_{\rm elementary} & =  i \bar f_{L i} \sla D  f_{L i} + i \bar f_{Ri} \sla D f_{Ri} ,  \\
\Lg_{\rm mixing} &= \l^{f_{L}}_{ij} \bar{f}_{Li} F_{Rj} + \l^{f_{R}}_{ij} \bar{F}'_{Li} f_{Rj} +{\rm h.c.} \, ,
\label{unifedcomposite}
\end{align}
where $F = (Q,L)$ and $F' =(U,D,N,E)$. 
The Lagrangian corresponds to 
the special case of Eqs.~(\ref{Lcomp0})-(\ref{Lmix}) with relations
\begin{align}
& \l^{q} = \l^{l} = \l^{f_{L}}, ~~~ M_{Q} = M_{L} = M_{F}, ~~~ Y^{U} = Y^{D} = Y^{N} = Y^{E}, \\
& \l^{u} = \l^{d} = \l^{\n} = \l^{e} = \l^{f_{R}}, ~~~ M_{U} = M_{D} = M_{N} = M_{E} = M_{F'}. 
\end{align}
If the gauge symmetry is not broken, it leads to the unrealistic GUT relation
\begin{align}
y^{f}_{\rm SM} &=  \l^{f_{L}} M_{F}^{-1} Y^{F} M_{F'}^{-1} \l^{f_{R}} .
\label{GUTyukawa}
\end{align}
Conversely, the GUT relation can be broken 
if the vev~(\ref{higgsvev}) contributes to the mass parameters. 

\subsection{Exotic fermions and their interactions}

In order to mediate GUT breaking flavor effects to the fermionic sector by renormalizable interactions, exotic fermions should be introduced. 
For this purpose, fields $\psi_{L}^{\a\b} ({\bf 6,1,1}) \sim (\tilde D^{c}_{R}, \tilde D_{L})$ and $\chi_{R}^{ab} ({\bf 1,2,2}) \sim (\tilde L^{c}_{L}, \tilde L_{R})$ appears to be a simple choice. 
They can form $\bf 10$ representation in $SO(10)$. 
Since they are real representations, the chiralities of fields $\psi_{L}$ and $\chi_{R}$ 
have no essential meaning. 
These exotic fermions are also assumed to couple to Higgs fields 
(and then would be globally charged in partially composite models). 

The general Yukawa interactions between 
these fields are written as
\begin{align}
\Lg &= 
Y^{6} \bar \psi_{L}^{\a\b} H_{R}^{\a a} F'_{R}{}^{\b a} 
+ Y^{22} \bar F_{L}^{\a a} H_{R}^{\a b} \chi_{R}^{ab} 
\\ &+ \tilde Y^{6} \bar F'_{L}{}^{\a a} H_{R}^{\dg \b a} \e^{\a\b\g\d} (\psi^{c}_{L})^{\g\d} 
+ \tilde Y^{22} \e^{ad} (\bar \chi^{c}_{R})^{de} \e^{eb}  H_{R}^{\dg \a a} F_{R}^{\a b}
+  {\rm h.c.} \, , 
\end{align}
By the vev~(\ref{higgsvev}), 
the exotics generate other linear mixing terms
\begin{align}
\Lg & \to 
Y^{6}_{ij} V \bar {\tilde D}_{L i}^{\a} D_{R j}^{\a} + Y^{22}_{ij} V \bar L_{L i}^{a} \tilde L_{R j}^{a} \\ 
&+ \tilde Y^{6}_{ij} V \bar D_{L i}^{\a} \tilde D_{R j}^{\a} 
+ \tilde Y^{22}_{ij} V \bar {\tilde L}_{L i}^{a} L_{R j}^{a}
+  {\rm h.c.} \, ,  
\end{align}
These exotics should have mass terms to mediate flavor structures.
Here we assume economical majorana mass
\begin{align}
\Lg &= {1\over 2} \tilde M_{\psi}  \bar \psi_{L}^{\a\b} \e^{\a\b\g\d} (\psi^{c}_{L})^{\g\d}  
+ {1\over 2} \tilde M_{\chi} \e^{ad} ( \bar \chi^{c}_{R})^{de} \e^{eb} \chi_{R}^{ab} 
+  {\rm h.c.}  \, , \\
&=  \tilde M_{\psi} \bar {\tilde D}_{L} \tilde D_{R}  
+ \tilde M_{\chi} \bar {\tilde L}_{L} \tilde L_{R}
+  {\rm h.c.}  \, .
\end{align}
Probably naive composite theories have vectorlike exotics $\psi_{L,R}, \chi_{L,R}$
with Dirac mass terms. 
However, extensions to the Dirac mass is trivial and
it has no influence on the later discussion. 

\subsection{Discussions of mass matrices}

Since the composites and exotics do not couple to the elementals, 
the mass matrix of down type quarks are extended to the following form 
\begin{align}
{\bar D_{L}}
\begin{pmatrix}
M_{F'} & \l^{f_{R}} \\
\end{pmatrix}
\column{D_{R}}{d_{R}}
 \to  
\row{\bar D_{L}}{\bar {\tilde D}_{L}}
\begin{pmatrix}
M_{F'} & \tilde Y^{6} V  & \l^{f_{R}} \\
Y^{6} V & M_{\psi} & 0
\end{pmatrix}
\Column{D_{R}}{\tilde D_{R}}{d_{R}} ,
\label{exmassmat}
\end{align}
and similar one holds for charged leptons. 

Decreases of the FN charges by fermion mixings require large mixings. 
General analysis is difficult because the model has so many free parameters and 
the diagonalization can not be done perturbatively. 
Nevertheless, several statements can be made from observations.  	
The lopsided texture in partially composite models 
is generated from the particular set of mass terms $\l^{f_{L}} M_{F}^{-1}$ and $M_{F'}^{-1} \l^{f_{R}}$. 
Then, decreases of FN charges are realized by
increases of $\l^{f_{L}, f_{R}}$ or decreases of $M_{F, F'}$. 
Since the couplings between exotics and elementals are forbidden, 
the increases of $\l^{f_{L}, f_{R}}$ from mixings are difficult without fine-tunings for $M_{F, F'}, Y^{6} V \gg \l^{f_{L}, f_{R}}$. 

Let us consider precisely the latter case, decreases of $M_{F, F'}$.
The later discussion of majorana neutrino mass suggests 
that the composite sector should have trivial flavor structures
 $Y^{6} V \propto \tilde Y^{6} V \propto M_{F,F'} \propto M_{\psi}$.
For simplicity, $Y^{6} = \tilde Y^{6}$ is also assumed.
In order to decrease the mass eigenvalues $M_{F' i}$ of $D_{(L,R)i}$
 to $\l^{2} M_{F' i}$, the mass parameters should satisfy the relation
$M_{F' i} \, M_{\psi i} \simeq Y^{6}_{ii} V \, \tilde Y^{6}_{ii} V$.
Treating the small mass eigenvalues $\l^{2} M_{F' i}$ as a perturbation, 
these parameters should satisfy the following relations
for the first and second generations:
\begin{align}
Y^{6}_{ii} V & \simeq \sqrt {M_{F' i} M_{\psi i}} (1 + \l^{2} X_{i}) 
\simeq \sqrt {M_{F' i} M_{\psi i}} \left(1 + { \l^{2}  (M_{F' i} + M_{\psi i}) \over 2 M_{\psi i}} \right) .
\end{align}
Here, the term $\l^{2} X_{i}$ is a perturbative correction to $Y^{6}_{ii} V$. 
These particular forms of Yukawa matrices $Y^{6}, \tilde Y^{6}$ (and $Y^{22}, \tilde Y^{22}$) produce the decreases of composite mass $M_{D,L}$
and the FN charges of $d_{R}, l_{L}$. 
However, it  requires 
an order of $\l^{2} \simeq 5 \%$ fine-tunings between 
matrices $Y^{6} V, \tilde Y^{6} V$ and $M_{F, F'}$.

As a result of the partial compositeness, 
the decreases of FN charges require
fine-tunings between mass and Yukawa matrices, 
either for the increases of $\l^{f,f'}$ or for the decreases of $M_{F,F'}$. 
Therefore, the case for $n=2$ and $n_{di} = n_{li} = (3,2,2)$, 
which requires only increases of FN charges 
will be appropriate to build a natural model. 

\subsection{Majorana Neutrino Mass}

The lepton number violation with in partially composite models have been discussed in literatures \cite{Agashe:2008fe, KerenZur:2012fr, Redi:2013pga}. 
In Refs.~\cite{Agashe:2008fe, KerenZur:2012fr} the majorana mass was introduced by an interaction between massive lepton and SM Higgs doublet $\bar L^{c}_{L} L_{L} HH / \L$.
On the other hand, Ref.~\cite{Redi:2013pga} introduced a majorana mass for elemental right-handed neutrinos, $M_{R \n} \bar \n_{R}^{c} \n_{R}$. 
In models with the type-I seesaw, 
the large mixing of light neutrinos can not be realized unless 
the hierarchy of $Y_{\n}$ and $M_{R}$ is compensated.
Then flavor structures $Y_{\n}$ and $M_{R}$ should be induced from the same mechanism. 

Here, we consider the other case, 
in which the massive leptons receive majorana mass terms from the GUT Higgs.
For this purpose, 
an exotic singlet $\s_{L} (1,1,1)$ are assumed. The general interactions between composites and the singlet are written as
\begin{align}
\Lg & =  {M_{\s} \over 2} \bar \s_{L} \s_{L}^{c} 
+ Y^{1 L} \bar \s_{L} H_{R}^{\dg \a a} F_{R}^{\a a} 
+ Y^{1 R} \bar F_{L}^{\a a} H_{R}^{\a a} \s_{L}^{c} 
+  {\rm h.c.}  \, . 
\end{align}
The mass matrix of them is found to be 
\begin{align}
\Row{\bar N_{L}}{\bar N_{R}^{c}}{\bar \s_{L}}
{1 \over 2}
\begin{pmatrix}
0 & M_{F'} & Y^{1R} V \\
M_{F'} & 0 &  (Y^{1L})^{T} V  \\
(Y^{1R})^{T}V  & Y^{1 L} V & M_{\s} 
\end{pmatrix}
\Column{N_{L}^{c}}{N_{R}}{\s_{L}^{c}}
+  {\rm h.c.}   \, .
\label{Nsmass}
\end{align}
The inclusive analysis is also difficult by the condition $M_{\rm GUT} \gtrsim M_{F'}$, which is required from decrease of the FN charges.
Although a specific diagonalization can not be described, 
the composite neutrinos $N_{(L,R)i}$ obtain majorana masses 
 $m_{(L,R) ij} \sim M_{F'}$.

When the mass matrix~(\ref{Nsmass}) is block diagonalized, 
the Lagrangian relevant to the majorana mass of light neutrinos is found to be
\begin{align}
\Lg & \ni  \z^{f_{R}}_{kj} \bar N_{Li} T_{ik} \n_{R j}  - \bar N_{L i} M_{Ni} N_{R i} 
- {1\over 2} m_{L ij} \bar N_{L i}^{c} N_{L j} - {1\over 2} m_{R ij} \bar N_{R i}^{c} N_{R j} 
+{\rm h.c.} \,.
\label{LNVL}
\end{align}
Here, $T_{ij}$ is some linear transformation, 
and the composite neutrino masses $M_{Ni}$ are generally different from the GUT invariant mass $M_{F' i}$. Due to this, the FN charge of the neutrinos $n_{\n i}$ can be changed from $(3,2,0)$. 

In the low energy, massive fields $N_{L,R}$ should be integrated out by using their equation of motions:
\begin{align}
{\del \Lg \over \del \bar N_{R i}^{c}} = 0
~ & \To ~ -M_{Ni}^{-1} m_{R ij} N_{R j}  =  N_{L i}^{c}, \\
{\del \Lg \over \del \bar N_{L i}} = 0 
~ &\To ~ T_{ik} \z^{f_{R}}_{kj}  \n_{R j} = 
{- m_{L ik} M_{Nk}^{-1} m_{R kj} N_{R j}} + M_{Ni} N_{R i} .
\label{firstterm} 
\end{align}
The first term in Eq.~(\ref{firstterm}) has roughly the same order
 as the second one. 
If this term can be neglected in some reason (c.f., $m_{L} = 0$), 
we can obtain a simple formula of $M_{\n R}$ similar to the Yukawa interactions~(\ref{SMYukawas}):
\begin{align}
\Lg \ni - {1\over 2} M_{\n R ij} \bar \n_{R i}^{c} \n_{R j} + {\rm h.c.} \, , ~~~
M_{\n R ij} = (\z'{}^{f_{R} T} M_{N}^{-1} m_{R} M_{N}^{-1} \z'{}^{f_{R}} )_{ij} ,
\end{align}
where $\z'{}^{f R} \equiv T \z^{f_{R}}.$

Therefore, the light neutrino mass can be calculated from the seesaw formula 
by integrated out the neutrinos $\n_{R i}$, 
\begin{align}
m_{\n}  &= {v^{2} \over 2} y_{\n} M_{\n R}^{-1} y_{\n}^{T} \\
&= {v^{2} \over 2}
(\z^{f_{L}} M_{F}^{-1} Y^{F} M_{N}^{-1} \z'{}^{f_{R}} )
( \z'{}^{f_{R} T} M_{N}^{-1} m_{R} M_{N}^{-1} \z'{}^{f_{R}} )^{-1}
( \z^{f_{L}} M_{F}^{-1} Y^{F} M_{N}^{-1} \z'{}^{f_{R}} )^{T} \\
&= {v^{2} \over 2}
(\z^{f_{L}} M_{F}^{-1} Y^{F}  m_{R}^{-1}  Y^{F T} M_{F}^{-1} \z^{f_{L} T} ) . 
\end{align}
In order to reproduce the lopsided texture, 
the flavor structure should be  generated from $\z^{f_{L}} M_{F}^{-1}$. 
In other words, $Y^{F} m_{R}^{-1} Y^{F T}$ and $Y^{F}$ should have (almost) the same flavor texture and then $m_{R}^{-1} Y^{F T} \simeq 1_{3}$. 
The lepton number violating mass $m_{R}$ 
is generated from $M_{F'}, Y^{1(L,R)}$ and $M_{\s}$ in Eq.~(\ref{Nsmass}). 
It means that 
composite neutrino sector should have (almost) the same flavor structure
 (c.f, $Y^{F} \propto M_{F'} \propto M_{\s} \propto Y^{1 (L,R)}$) to reproduce the large mixing of neutrinos by the seesaw formula.
If the first term in Eq.~(\ref{firstterm}) can not be neglected, 
the same discussion can be applied for $m_{L}$. 
For example, these conditions can be realized 
by a $SU(3)_{F}$ flavor symmetry in the composite sector. 

\section{Conclusions and Discussion}

In this paper, we consider a lopsided flavor texture compatible with thermal leptogenesis 
in partially composite Pati--Salam unification. 
The Davidson--Ibarra bound $M_{\nu R1} \gtrsim 10^9 \GeV$ for the successful thermal leptogenesis can be recast to the Froggatt--Nielsen (FN) charge of the lopsided texture. 
We found the FN charge $n_{\n1}$ of the lightest right-handed neutrino $\n_{R1}$ can not be larger than a upper bound, $n_{\n1} \lesssim 4.5$.

From the viewpoint of unification, the FN charges of the neutrinos $n_{\n i}$ 
should be the same to that of other SM fermions. 
Then, two cases $n_{\n i} = n_{qi} = (3,2,0)$ and $ n_{\n i} = n_{l i} = (n+1,n,n)$ are considered. 
Observations of PS model shows that the case of $n=0$, 
$n_{li} = n_{di} = (1,0,0)$ will be the simplest realization. 

To decrease the FN charges of these fermions from the GUT invariant FN charges
$n_{qi} = (3,2,0)$, we utilize the partial compositeness. 
In this picture, the hierarchies of Yukawa matrices are a consequence of 
mixings between massless chiral fermions $f_{L}, f'_{R}$ and massive vector fermions $F_{L,R}, F'_{L,R}$. 
This is induced by the linear mixing terms $\l^{f} \bar f_{L} F_{R}$ and $\l^{f'} \bar F'_{L} f'_{R}$. 

The lopsided texture in partially composite models 
is generated from the particular set of mass terms $\l^{f_{L}} M_{F}^{-1}$ and $M_{F'}^{-1} \l^{f_{R}}$. Then, decreases of FN charges are realized by 
increases of $\l^{f_{L}, f_{R}}$ or decreases of $M_{F, F'}$. 
Since the couplings between exotics and elementals are forbidden, 
the increases of $\l^{f_{L}, f_{R}}$ from mixings are difficult without fine-tunings for $M_{F, F'}, Y^{6} V \gg \l^{f_{L}, f_{R}}$. 
On the other hand, the decreases of $M_{F,F'}$ require an order of $\l^{2} \simeq 5 \%$ fine-tunings between matrices $Y^{6} V, \tilde Y^{6} V$ and $M_{F, F'}$. 

As a result of the partial compositeness, 
the decreases of FN charges require
fine-tunings between mass and Yukawa matrices 
either for the increases of $\l^{f, f'}$ or for the decreases of $M_{F,F'}$. 
Therefore, the case for $n=2$ and $n_{di} = n_{li} = (3,2,2)$, 
which requires only increases of FN charges 
will be appropriate to build a natural model. 

Moreover, it is found that 
composite neutrino sector should have (almost) the same flavor structure
 to reproduce the large mixing of neutrinos by the type-I seesaw mechanism.
If the vev of GUT breaking Higgs mediates flavor structure, 
they contribute to some mass term.  
Then, this statement  can be hold for even in other Pati--Salam model, 
that does not assume the partial compositeness.  

\section*{Acknowledgement}

The author would like to appreciate R. Kitano for useful discussions of partial compositeness, 
and Y. Takanishi for valuable comments of leptogenesis.
This study is financially supported by the Iwanami Fujukai Foundation.


\end{document}